\documentclass[peerreview, 11pt]{IEEEtran}
\usepackage[dvips]{graphicx,epsfig}
\usepackage{epsfig}
\usepackage{graphicx}
\usepackage{subfigure}

\usepackage{amsmath}
\usepackage{amssymb}
\usepackage{color}
\usepackage{colordvi}
\usepackage{float}

\begin{document}

\setlength{\leftmargini}{0\leftmargini}
\newtheorem{definition}{Definition}
\newtheorem{lemma}{Lemma}
\newtheorem{proposition}{Proposition}
\newtheorem{corollary}{Corollary}
\newtheorem{theorem}{Theorem}
\newtheorem{conjecture}{Conjecture}
\newtheorem{remark}{Remark}

\newcommand{\dsum}{\displaystyle\sum}
\newcommand{\dfr}{\displaystyle\frac}
\newcommand{\dint}{\displaystyle\int}
\newcommand{\dprod}{\displaystyle\prod}
\newcommand{\naturals}{\ensuremath{\mathbb{N}}}
\newcommand{\reals}{\ensuremath{\mathbb{R}}}
\newcommand{\expectation}{\ensuremath{\mathbb{E}}}

\title{\bf{Upper Bound on Error Exponent of Regular LDPC Codes Transmitted over the BEC}}


\author{Idan Goldenberg \hspace*{1cm} David Burshtein \\
School of Electrical Engineering \\
Tel-Aviv University \\
Tel-Aviv 69978, Israel \\
E-mail: {\tt \{idang,burstyn\}@eng.tau.ac.il} }

\date{}
\maketitle

\baselineskip=18pt
\begin{abstract}
\baselineskip=16pt The error performance of the ensemble of typical
LDPC codes transmitted over the binary erasure channel (BEC) is
analyzed. In the past, lower bounds on the error exponents were
derived. In this paper a probabilistic upper bound on this error
exponent is derived. This bound holds with some confidence level.
\end{abstract}


\section{Introduction}
Low-density parity-check (LDPC) codes, discovered by Gallager
\cite{Gallager1963}, have been widely researched over the last
decade and a half. Asymptotic results are widely known for these
codes, including results on the performance under maximum-likelihood
(ML) decoding
\cite{Gallager1963,MacKay99,Montanari2001,Burshtein_Miller_IT01,BarakBurshtein_2005},
average ensemble distance spectra
\cite{Gallager1963,LitsynShevelev2002,Burshtein_Miller_IT04,Rathi2006,DiRiU2006},
stopping set distributions
\cite{Burshtein_Miller_IT04,Rathi2006,DiRiU2006,Orlitsky2005},
thresholds for iterative decoding using density evolution
\cite{LMSS2001,RichardsonUrbanke2001DE}, and others. However,
accurate finite-length analysis of LDPC codes under iterative
sum-product decoding is currently available only for the binary
erasure channel (BEC) \cite{DiProRiTelU}. This is due to the
simplicity of the channel model and the graph-based iterative
decoder which lends itself to a more detailed analysis. Analysis of
the combinatorial properties of stopping sets and their contribution
to the error performance reveals that the average error performance
of the LDPC ensemble is proportional to the inverse of a polynomial
in the block length $N$ \cite{Burshtein_Miller_IT04}. This behavior
is attributed to the existence of ``bad'' codes which possess small
stopping sets, and otherwise would decrease \emph{exponentially}
with $N$ if these codes were removed from the ensemble. Fortunately,
these ``bad'' codes constitute a small fraction of the entire
ensemble whose size is proportional to the inverse of a polynomial
in $N$.

After removing the undesirable codes, we obtain an \emph{expurgated
ensemble}, for which there exists a positive error exponent. In
\cite{Burshtein_Miller_IT04}, lower bounds on this error exponent of
typical codes in the regular and irregular LDPC code ensembles were
derived. In this paper we obtain an upper bound on this exponent,
and compare it with the above mentioned lower bounds. Similar to
\cite{BarakBurshtein_2005}, which considers upper bounds on the
error exponent of LDPC codes under ML decoding, our bounds depend on
some confidence level.

The correspondence is organized as follows. Section
\ref{Preliminaries section} introduces notation and preliminary
material. Section \ref{Upper Bound on Error Exponent for the BEC}
introduces a lower bound on the error (erasure) probability from
which an upper bound on the exponent is derived. Section
\ref{Numerical Results} introduces numerical results and
comparisons with previous results. Section \ref{Conclusion}
concludes the paper.

\section{Preliminaries}\label{Preliminaries section}

\subsection{Notation}
We will use the following notation throughout the paper.
\begin{itemize}
\item Let $\{\alpha_l \}_{l=1}^k$ be a set of non-negative real
numbers, such that $\sum_l \alpha_l \leq 1$. The entropy function
of $\{\alpha_l \}_{l=1}^k$ is defined as

\begin{equation*}
    h\left(\alpha_1,\ldots,\alpha_k\right)=-\sum_{l=1}^k \alpha_l
\log(\alpha_l) - \left(1-\sum_{l=1}^k
\alpha_l\right)\log\left(1-\sum_{l=1}^k \alpha_l\right)
\end{equation*}
where $\log$ is the base-2 logarithm. We use the convention
$0\log0 =0$.

\item
 Given an integer $n$ and integers $(n_1,\ldots,n_k)$ such that
$\sum_l n_l \leq n$,
\begin{equation*}\label{multinomial coefficient definition}
    \binom{n}{n_1,n_2,\ldots,n_k} \triangleq \frac{n!}{n_1! \cdot n_2! \cdot \ldots \cdot \left( n - \sum_{l=1}^k n_l \right) !}
\end{equation*}
is the multinomial coefficient of $n$ over $(n_1,\ldots,n_k)$. We
will use the following property of multinomial coefficients
\begin{equation}\label{Multinomial coeffieient property}
    \log \binom{n}{n_1,n_2,\ldots,n_k} = n \left(
h\left(\frac{n_1}{n},\ldots,\frac{n_k}{n}\right) + o(1) \right)
\end{equation}
which is easily proven using Stirling's approximation.

\item
    If $p(x)$ is a polynomial, then we will denote the coefficient of $x^i$ by
$\left[x^i\right] p(x)$, i.e,
\begin{equation*}
    p(x)=\sum_i \left[x^i\right] p(x) x^i
\end{equation*}
The same notation is extended for use with multivariate
polynomials, e.g.,
\begin{equation*}
    p(x,y,z)=\sum_{i,j,k} \left[x^i y^j z^k \right] p(x,y,z) x^i y^j
z^k
\end{equation*}

\end{itemize}

\subsection{A Second-Order Inequality for Probabilities}
Dawson and Sankoff \cite{DawsonSankoff1967} obtained a lower bound
on the probability of a finite union of events. Their result
asserts the following. Let $\{A_i\}_{i=1}^M$ be a finite family of
events in a probability space $(\Omega,P)$. Denote
\begin{equation*}
    \tilde{S}_1 = \sum_{i \in I}\Pr (A_i) \quad \quad \tilde{S}_2=\sum_{\substack{i,j \in I \\ i > j}}\Pr (A_i \cap A_j)
\end{equation*}
where $I=\{1,\dots,M\}$. Then
\begin{equation}\label{Dawson's inequality}
    \Pr \left( \bigcup_{i \in I}A_i \right) \geq \frac{2}{r+1} \tilde{S}_1 - \frac{2}{r(r+1)}
\tilde{S}_2
\end{equation}
for any $r \in \{1,\dots,M-1\}$.

Following the derivation in \cite{DawsonSankoff1967}, we derive a
result which generalizes \eqref{Dawson's inequality}. For a
probability event $A$, denote by $\mathbf{1}_{\{A\}}$ to be the
indicator (random variable) over $A$, i.e, for $\omega \in
\Omega$,
\begin{equation*}\label{indicator}
    \mathbf{1}_{\{A\}}(\omega) = \left\{ \begin{array}{l l} 1 &
\omega \in A \\ 0 & \omega \notin A \end{array} \right.
\end{equation*}

Our result asserts that for all $\omega \in \Omega$,

\begin{equation}\label{Dawson's generalized inequality}
    \mathbf{1}_{\left\{ \cup_{i=1}^M A_i \right\}} \geq
\frac{2}{r+1} S_1 - \frac{2}{r(r+1)} S_2
\end{equation}
where
\begin{equation*}
    S_1 = \sum_{i \in I} \mathbf{1}_{\{A_i\}} \quad \quad S_2=\sum_{\substack{i,j \in I \\ i > j}}
\mathbf{1}_{\{A_i\}}\mathbf{1}_{\{A_j\}}
\end{equation*}
By taking the expectation over both sides of \eqref{Dawson's
generalized inequality}, we get \eqref{Dawson's inequality} as a
special case. We prove \eqref{Dawson's generalized inequality} in
Appendix \ref{Proof of Dawson's generalized}.

\subsection{LDPC Code Ensembles}
We consider the standard bipartite graph-based $(c,d)$-regular
LDPC code ensemble with block length $N$ and design rate $R$. In
this ensemble a randomly chosen permutation is used to match the
$cN$ left sockets to the $d(1-R)N$ right sockets. The actual rate
of the code is at least $R \triangleq 1-c/d$.

\section{Upper Bound on Error Exponent for the BEC}\label{Upper Bound on Error Exponent for the BEC}
Recall that a \emph{stopping set} $\mathcal{S}$ of a bipartite graph
representation of an LDPC code is a set of variable nodes, such that
each check node neighbor of $\mathcal{S}$ is connected to
$\mathcal{S}$ by at least two edges. As explained in
\cite{DiProRiTelU}, iterative decoding of LDPC codes succeeds if and
only if the set of variable nodes which correspond to erasures does
not contain a subset which is a stopping set.

 The \emph{expurgated} $(c,d)$-regular LDPC ensemble
$\mathcal{C}^{\gamma}$ is derived from the $(c,d)$-regular
ensemble $\mathcal{C}^0$ by removing all the codes containing
stopping sets of size $\gamma N$ or less. It was shown in
\cite{Burshtein_Miller_IT04} that for ensembles with $c>2$, if
$\gamma$ is selected below a certain threshold $\alpha_0$, then
almost all codes in $\mathcal{C}^0$ belong to
$\mathcal{C}^{\gamma}$. In other words, if $\mathcal{C}$ is drawn
at random from $\mathcal{C}^0$
\begin{equation}\label{expurgated ensemble is almost equal to regular}
    \Pr \left( \mathcal{C} \in \mathcal{C}^{\gamma} \right) =
    1-o(1) \quad \forall \gamma<\alpha_0
\end{equation}
The number $\alpha_0 N$ may therefore be considered to be the
typical minimum stopping set size of $\mathcal{C}^0$. Since the
behavior of $\mathcal{C}^0$ is dominated by a small fraction of
``bad'' codes, we will be interested in the performance of codes
drawn at random from $\mathcal{C}^{\gamma}$. Let $\mathcal{C}$ be
such a code.

Consider a BEC with erasure probability $\delta$; the probability
of unsuccessful decoding of any codeword from $\mathcal{C}$,
$P_e^{\mathcal{C}}$ is given by
\begin{equation}\label{Probability of erasure}
    P_e^{\mathcal{C}} = \sum_{l=\gamma N}^N  \delta^l
(1-\delta)^{N-l} \sum_m \mathbf{1}_{\left\{\cup_{i=1}^{2^l-1}
A_i^m\right\}}
\end{equation}
where the index $m$ runs over all sets of variable nodes containing
exactly $l$ nodes; for a particular set $\mathcal{S}_m$ of $l$
variable nodes, $\{A_i^m\}$ is the event that the $i$'th (non-empty)
subset of $\mathcal{S}_m$ (where $i=1,\dots,2^l-1$) is a stopping
set. Note that every set of $N(1-R)+1$ variable nodes contains the
support of a nonzero codeword\footnote{This is tantamount to saying
that $N(1-R)+1$ columns in the parity check matrix, regardless of
how they are chosen, are linearly dependent; this follows since the
matrix has $N(1-R)$ rows.}. Hence (since every codeword is a
stopping set), every set of $N(1-R)+1$ variable nodes contains a
stopping set. Therefore, the indicator appearing in the RHS of
\eqref{Probability of erasure} may be replaced by $1$ for
$l>N(1-R)$, which yields
\begin{equation}\label{Probability of erasure, true indicator}
    P_e^{\mathcal{C}} = \sum_{l=\gamma N}^{N(1-R)}  \delta^l
(1-\delta)^{N-l} \sum_m \mathbf{1}_{\left\{\cup_{i=1}^{2^l-1}
A_i^m\right\}} + \sum_{l= N(1-R)+1}^{N} \binom{N}{l} \delta^l
(1-\delta)^{N-l}
\end{equation}
Next, we use \eqref{Dawson's generalized inequality} to
lower-bound the indicator function in \eqref{Probability of
erasure, true indicator}, giving
\begin{equation}\label{Application of Dawson's generalized inequality}
    \mathbf{1}_{\left\{ \cup_{i=1}^{2^l-1} A_i^m \right\}} \geq
\frac{2}{r_l+1} S_1 - \frac{2}{r_l(r_l+1)} S_2
\end{equation}
where $r$ is allowed to depend on the size of the set, and
\begin{equation}\label{S_1 and S_2}
    S_1 = \sum_{i=1}^{2^l-1} \mathbf{1}_{\{A_i^m\}} \quad \quad
S_2=\sum_{i=1}^{2^l-1}\sum_{k=1}^{i-1}
\mathbf{1}_{\{A_i^m\}}\mathbf{1}_{\{A_k^m\}}
\end{equation}
Consider a stopping set $\mathcal{S}$ containing $k$ variable
nodes, where $k\leq l$. The number of sets of variable nodes of
size $l$ containing $\mathcal{S}$ as a subset is
$\binom{N-k}{l-k}$. Hence, again letting $m$ run over all subsets
of size $l$, we have
\begin{equation}\label{getting j out of S1}
    \sum_m \sum_{i=1}^{2^l-1} \mathbf{1}_{\{A_i^m\}} =
\sum_{k=1}^l \binom{N-k}{l-k} S_k^{\mathcal{C}} = \sum_{k=\gamma
N}^l \binom{N-k}{l-k} S_k^{\mathcal{C}}
\end{equation}
where $S_k^{\mathcal{C}}$ is the number of stopping sets with $k$
variable nodes in $\mathcal{C}$; note that since $\mathcal{C}$
belongs to the expurgated ensemble, we have $S_k^{\mathcal{C}}=0$
for $k<\gamma N$.

In a similar fashion we obtain
\begin{equation}\label{getting j out of S2}
    \sum_m
\sum_{i=1}^{2^l-1}\sum_{j=1}^{i-1}\mathbf{1}_{\{A_i^m\}}
\mathbf{1}_{\{A_j^m\}} = \sum_{\substack{\gamma N \leq j \leq i \leq l \\
 0 \leq k \leq j+\min(i-j-1,0) \\ i+j-k \leq l}}
\binom{N-(i+j-k)}{l-(i+j-k)}S_{i,j,k}^{\mathcal{C}}
\end{equation}
where $S_{i,j,k}^{\mathcal{C}}$ is the number of \emph{pairs} of
stopping sets, $\left( \mathcal{S}_1,\mathcal{S}_2\right)$
satisfying $|\mathcal{S}_1|=i$, $|\mathcal{S}_2|=j$, and
$|\mathcal{S}_1 \cap \mathcal{S}_2|=k$. Recalling that both
$\mathcal{S}_1$ and $\mathcal{S}_2$ must be subsets of a particular
set of size $l$, their union must also be a subset, and therefore
$|\mathcal{S}_1 \cup \mathcal{S}_2|=i+j-k \leq l$. Furthermore, the
application of \eqref{Dawson's generalized inequality} requires
summing over pairs of \emph{distinct} events. Consequently, we
cannot have $\mathcal{S}_1 = \mathcal{S}_2$, i.e., when $i=j$, we
must have $k<j$; this requirement is subsumed by imposing $0 \leq k
\leq j+\min(i-j-1,0)$ in \eqref{getting j out of S2}. Plugging
\eqref{Application of Dawson's generalized
inequality}-\eqref{getting j out of S2} into \eqref{Probability of
erasure, true indicator}, we get
\begin{eqnarray*}
  P_e^{\mathcal{C}} &\geq& \sum_{l=\gamma N}^{N(1-R)} \delta^l (1-\delta)^{N-l}
\left[\frac{2}{r_l+1}\sum_{i'=\gamma N}^l
\binom{N-i'}{l-i'}S_{i'}^{\mathcal{C}}  \right. \nonumber \\
&& \left. -\frac{2}{r_l(r_l+1)}
\sum_{\substack{\gamma N \leq j \leq i \leq l \\
 0 \leq k \leq j+\min(i-j-1,0) \\ i+j-k \leq l}}
\binom{N-(i+j-k)}{l-(i+j-k)}S_{i,j,k}^{\mathcal{C}} \right]
+ \sum_{l=N(1-R)+1}^N \binom{N}{l}\delta^l (1-\delta)^{N-l} \nonumber \\
&\geq& \sum_{l=\gamma N}^{N(1-R)} \left\{\delta^{N\epsilon}
(1-\delta)^{N(1-\epsilon)} \left[\frac{2}{r_l+1} \max_{\gamma \leq
\eta \leq \epsilon} \binom{N(1-\eta)}{N(\epsilon-\eta)} S_{\eta
N}^{\mathcal{C}} \right. \right. \nonumber \\ && \left. \left.
-\frac{2}{r_l(r_l+1)}{(\epsilon N)}^3 \max_{\substack{\gamma \leq
\eta_2 \leq \eta_1 \leq \epsilon \\ 0 \leq \beta \leq \eta_2
\\ \eta_1 + \eta_2 -\beta \leq \epsilon}} \binom{N(1-(\eta_1 + \eta_2 -\beta))}
{N(\epsilon-(\eta_1 + \eta_2 -\beta))}S_{\eta_1 N,\eta_2 N,\beta
N}^{\mathcal{C}} \right] \right\} \nonumber \\ && + \max_{1-R \leq
\epsilon \leq 1 } \left\{ \binom{N}{N\epsilon}
\delta^{N\epsilon}(1-\delta)^{N(1-\epsilon)} \right\}
\nonumber \\
&\stackrel{(a)}{\geq}& \max_{\gamma \leq \epsilon \leq 1-R}
\left\{\delta^{N\epsilon} (1-\delta)^{N(1-\epsilon)}
\hat{P}_e^{\mathcal{C}}(\epsilon,N) \right\} + \max_{1-R \leq
\epsilon \leq 1 } \left\{ \binom{N}{N\epsilon}
\delta^{N\epsilon}(1-\delta)^{N(1-\epsilon)} \right\}
\end{eqnarray*}
where
\begin{eqnarray}\label{PeCtag}
    \hat{P}_e^{\mathcal{C}}(\epsilon,N) &\triangleq& \left[\frac{2}{r_{\epsilon N}+1} \max_{\gamma \leq \eta \leq \epsilon}
\binom{N(1-\eta)}{N(\epsilon-\eta)} S_{\eta N}^{\mathcal{C}} \right.
\nonumber \\ && \left.  -\frac{2}{r_{\epsilon N}(r_{\epsilon
N}+1)}{(\epsilon N)}^3 \max_{\substack{\gamma \leq \eta_2 \leq
\eta_1 \leq \epsilon
\\ 0 \leq \beta \leq \eta_2
\\ \eta_1 + \eta_2 -\beta \leq \epsilon}} \binom{N(1-(\eta_1 + \eta_2 -\beta))}
{N(\epsilon-(\eta_1 + \eta_2 -\beta))}S_{\eta_1 N,\eta_2 N,\beta
N}^{\mathcal{C}} \right]
\end{eqnarray}
and $\epsilon \triangleq \frac{l}{N}$, $\eta \triangleq
\frac{i'}{N}$, $\eta_1 \triangleq \frac{i}{N}$, $\eta_2 \triangleq
\frac{j}{N}$, and $\beta \triangleq \frac{k}{N}$; a sufficient
condition in order for (a) to hold is that
$\hat{P}_e^{\mathcal{C}}(\epsilon,N)$ be non-negative for $\gamma
\leq \epsilon \leq 1-R$. Later we will choose the value of
$r_{\epsilon N}$ so that this condition is fulfilled.

By expressing the bound in exponential form, we get the following
upper bound on the error exponent
\begin{equation*}
    -\frac{1}{N}\log P_e^{\mathcal{C}} \leq -\max_{\gamma \leq \epsilon \leq
1} \left\{\epsilon \log \delta +(1-\epsilon) \log (1-\delta) +
\left\{ \begin{array}{l l} \frac{1}{N}\log P_e^{\mathcal{C}}
(\epsilon, N) & \gamma \leq \epsilon \leq 1-R \\
h(\epsilon) & 1-R \leq \epsilon \leq 1 \end{array} \right.
\right\} + o(1)
\end{equation*}
where we rely upon \eqref{Multinomial coeffieient property}, and
\begin{eqnarray}\label{Pex exponential}
    P_e^{\mathcal{C}}(\epsilon, N) &\triangleq& \frac{2}{r_{\epsilon N}+1}2^{-N
E'_1}-\frac{2}{r_{\epsilon N}(r_{\epsilon N}+1)}2^{-N E'_2} \label{Pe^C} \\
E'_1 &=& -\max_{\gamma \leq \eta \leq \epsilon} \left\{ (1-\eta)h
\left(\frac{\epsilon-\eta}{1-\eta} \right) + \frac{1}{N}\log S_{\eta
N}^{\mathcal{C}}  \right\} \label{E_1'} \\
E'_2 &=& -\max_{\substack{\gamma \leq \eta_2 \leq \eta_1 \leq
\epsilon
\\ 0 \leq \beta \leq \eta_2 \\ \eta_1 + \eta_2 -\beta \leq \epsilon}} \left\{ (1-(\eta_1 + \eta_2
-\beta)) h \left(\frac{\epsilon-(\eta_1 + \eta_2
-\beta)}{1-(\eta_1 + \eta_2 -\beta)}\right) + \frac{1}{N}\log
S_{\eta_1 N,\eta_2 N,\beta N}^{\mathcal{C}}  \right\} \label{E_2'}
\end{eqnarray}
Let $\mathcal{C'}$ be a randomly selected code from
$\mathcal{C}^0$, and let $\overline{S}_i$ and
$\overline{S}_{i,j,k}$ be the averages, over $\mathcal{C}^0$, of
$S_i^{\mathcal{C'}}$ and $S_{i,j,k}^{\mathcal{C'}}$, respectively.
We evaluate these average quantities and then relate them to
$S_{i}^{\mathcal{C}}$ and
$S_{i,j,k}^{\mathcal{C}}$\footnote{recall that in our context
$\mathcal{C}$ is selected uniformly from $\mathcal{C}^{\gamma}$}.
In order to evaluate these quantities, we introduce the following
notation.
\begin{eqnarray}
    \psi_i(x;d)&=&\sum_{l=i}^d
\binom{d}{l}x^l=(1+x)^d-\sum_{l=0}^{i-1} \binom{d}{l}x^l
\label{little psi}
\\ \Psi_{i_-,k_-,j_-}^{i_+,k_+,j_+}(x,y,z,d)&=&
\sum_{\substack{i_-\leq i \leq i_+\\j_-\leq j \leq j_+\\k_-\leq k
\leq k_+ \\i+j+k\leq d}} \binom{d}{i,j,k}x^i y^j z^k \label{big psi}
\end{eqnarray}
The average quantities satisfy
\begin{eqnarray}
  \overline{S}_i &=& \binom{N}{i}P_{s,1}(i) \label{Si}\\
  \overline{S}_{i,j,k} &=& \binom{N}{i-k,k,j-k}P_{s,2}(i,j,k)
  \label{Sijk}
\end{eqnarray}
where $P_{s,1}(i)$ is the probability that a specific set of
variable nodes, $\mathcal{S}$, is a stopping set, and
$P_{s,2}(i,j,k)$ is the probability that a specific pair of sets -
$\mathcal{S}_1$ containing $i$ variable nodes and $\mathcal{S}_2$
containing $j$ variable nodes, with $|\mathcal{S}_1 \cap
\mathcal{S}_2|=k$, are both stopping sets.

To evaluate $P_{s,1}(i)$, we need to fix a set $\mathcal{S}$ of
$i$ variable nodes and count the number of possibilities of
connecting their $ic$ variable sockets to $ic$ check sockets such
that each of the $L$ check nodes is either (a) not connected to
any of the $ic$ variable sockets, or (b) connected by at least two
check sockets. This combinatorial problem can be solved by means
of the enumeration function in \eqref{little psi}. The total
number of ways to connect $ic$ variable sockets to $Nc$ check
sockets is $\binom{Nc}{ic}$, therefore
\begin{equation*}\label{P_(s,1)(i)}
    P_{s,1}(i)=\frac{\left[ x^{ic}\right] \left(1+\psi_2(x,d)\right)^L}{\binom{Nc}{ic}}
\end{equation*}

\begin{figure}[hbt]
\begin{center}
\leavevmode
\input{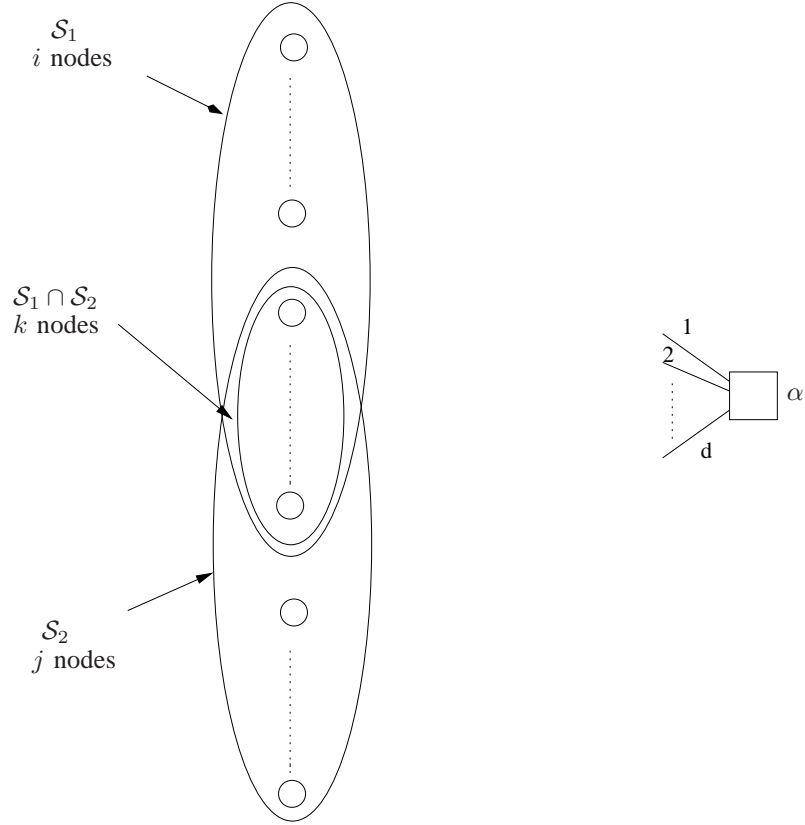}
\caption{Two intersecting stopping sets and a check node $\alpha$}
\label{Stopping set and intersection}
\end{center}
\end{figure}

We proceed with the evaluation of $P_{s,2}(i,j,k)$. Given two sets
$\mathcal{S}_1$ and $\mathcal{S}_2$ of variable nodes with
$|\mathcal{S}_1|=i$, $|\mathcal{S}_2|=j$, $|\mathcal{S}_1\cap
\mathcal{S}_2|=k$, we need to count the number of possibilities of
connecting $(i-k)c$ sockets from $\mathcal{S}_1/\mathcal{S}_2$,
$kc$ sockets from $\mathcal{S}_1\cap \mathcal{S}_2$ and $(j-k)c$
sockets from $\mathcal{S}_2/\mathcal{S}_1$ to $(i+j-k)c$ check
sockets, such that both $\mathcal{S}_1$ and $\mathcal{S}_2$ are
stopping sets. This situation is depicted in Figure \ref{Stopping
set and intersection}. Consider a check node $\alpha$ in the
graph. From the definition of a stopping set, it can be seen that
in order to have both $\mathcal{S}_1$ and $\mathcal{S}_2$ as
stopping sets, $\alpha$ has to fall into one of the following
disjoint categories:
\begin{itemize}
\item
    $\alpha$ is not connected at all to nodes in $\mathcal{S}_1\cup
\mathcal{S}_2$.

\item
    $\alpha$ is connected by at least two edges to nodes in
$\mathcal{S}_1/\mathcal{S}_2$ and is not connected to nodes in
$\mathcal{S}_2$.

\item
    $\alpha$ is connected by at least two edges to nodes in
$\mathcal{S}_2/\mathcal{S}_1$ and is not connected to nodes in
$\mathcal{S}_1$.

\item
    $\alpha$ is connected by at least two edges to nodes in
$\mathcal{S}_1/\mathcal{S}_2$ and by at least two edges to nodes
in $\mathcal{S}_2/\mathcal{S}_1$, but is not connected to any node
in $\mathcal{S}_1 \cap \mathcal{S}_2$.

\item
    $\alpha$ is connected by exactly one edge to a node in $\mathcal{S}_1 \cap
\mathcal{S}_2$, and by at least one edge to nodes in
$\mathcal{S}_1/\mathcal{S}_2$ and in
$\mathcal{S}_2/\mathcal{S}_1$.

\item
    $\alpha$ is connected by at least two edges to nodes in $\mathcal{S}_1 \cap
\mathcal{S}_2$.

\end{itemize}
This combinatorial problem can be solved using the enumeration
function given in \eqref{big psi}. The total number of
possibilities of connecting $(i-k)c$ sockets from
$\mathcal{S}_1/\mathcal{S}_2$, $kc$ sockets from
$\mathcal{S}_1\cap \mathcal{S}_2$ and $(j-k)c$ sockets from
$\mathcal{S}_2/\mathcal{S}_1$ to $Nc$ check sockets is
$\binom{Nc}{(i-k)c,kc,(j-k)c}$. Therefore,
\begin{eqnarray}\label{P_(s,2)(i)}
    P_{s,2}(i,j,k)&=&\left[x^{(i-k)c}y^{kc}z^{(j-k)c}\right] B(x,y,z,d)^L \cdot
\binom{Nc}{(i-k)c,kc,(j-k)c}^{-1} \nonumber \\
B(x,y,z,d)&\triangleq&1 +
\Psi_{2,0,0}^{d,0,0}(x,y,z,d)+\Psi_{0,0,2}^{0,0,d}(x,y,z,d) +
\Psi_{2,0,2}^{d-2,0,d-2}(x,y,z,d) \nonumber \\ && +
\Psi_{1,1,1}^{d-1,1,d-1}(x,y,z,d) + \Psi_{0,2,0}^{d,d,d}(x,y,z,d)
\end{eqnarray}

We turn our attention back to the relation between the average
quantities $\overline{S}_i$ and $\overline{S}_{i,j,k}$ and those of
the randomly selected code, $S_{i}^{\mathcal{C}}$ and
$S_{i,j,k}^{\mathcal{C}}$. By assuming that $\mathcal{C}$ is
selected at random with uniform probability from $\mathcal{C}^0$ and
using conditioning, we have
\begin{eqnarray}
  \Pr \left( S_{i,j,k}^{\mathcal{C}} > N \overline{S}_{i,j,k} \; | \; \mathcal{C} \in \mathcal{C}^{\gamma}
   \right) &=& \frac{\Pr \left( S_{i,j,k}^{\mathcal{C}} > N \overline{S}_{i,j,k} \right)
   - \Pr \left(\mathcal{C} \notin \mathcal{C}^{\gamma},S_{i,j,k}^{\mathcal{C}} > N \overline{S}_{i,j,k} \right) }
   {\Pr \left( \mathcal{C} \in \mathcal{C}^{\gamma} \right)} \nonumber \\
  &\stackrel{\mathrm{(a)}}{\leq}& \frac{\Pr \left( S_{i,j,k}^{\mathcal{C}} > N
  \overline{S}_{i,j,k} \right)}{1-o(1)} \stackrel{\mathrm{(b)}}{\leq}
  \frac{1}{N(1-o(1))} \label{inequality for E_2}
\end{eqnarray}
where (a) is obtained using \eqref{expurgated ensemble is almost
equal to regular} and by omitting the negative term, and (b) is due
to Markov's inequality. We conclude from \eqref{inequality for E_2}
that w.p. (with probability) $1 - o(1)$, for $\mathcal{C}$ chosen
randomly with uniform probability from $\mathcal{C}^{\gamma}$,
\begin{equation}\label{inequality for Sijk}
    \frac{1}{N} \log S_{i,j,k}^{\mathcal{C}} \leq \frac{1}{N} \log
    \overline{S}_{i,j,k} + o(1)
\end{equation}
By using conditioning once more we obtain
\begin{eqnarray}
  \Pr \left( 1-\epsilon \leq \frac{S_{i}^{\mathcal{C}}}
  {\overline{S}_{i}} \leq 1+\epsilon \; \Big| \; \mathcal{C} \in \mathcal{C}^{\gamma} \right)
  &\geq& \frac{\Pr \left( 1-\epsilon \leq \frac{S_{i}^{\mathcal{C}}}
  {\overline{S}_{i}} \leq 1+\epsilon \right)-\Pr \left( \mathcal{C} \notin \mathcal{C}^{\gamma} \right)}
  {\Pr \left( \mathcal{C} \in \mathcal{C}^{\gamma} \right)} \nonumber \\
  &\stackrel{\mathrm{(a)}}{\geq}& \Pr \left( 1-\epsilon \leq \frac{S_{i}^{\mathcal{C}}}
  {\overline{S}_{i}} \leq 1+\epsilon \right) +o(1)
  \label{Conditioning on Si}
\end{eqnarray}
where (a) is obtained by using \eqref{expurgated ensemble is
almost equal to regular} and replacing the denominator by $1$.

Rathi \cite{Rathi2006} has obtained a concentration result on the
stopping set distribution. His result implies the following. For any
$\epsilon>0$,
\begin{equation}\label{Rathi}
    \Pr \left( 1-\epsilon \leq \frac{S_{\eta N}^{\mathcal{C}}}{\overline{S}_{\eta N}} \leq 1+
    \epsilon
    \right) \geq 1 - \frac{\beta_{\eta,d,c}}{\epsilon^2} + o(1)
\end{equation}
where $\beta_{\eta,d,c}$ is a constant given in Eq. \eqref{rathi's
result} in Appendix~\ref{Confidence Interval of Stopping Set
Distribution}, independent of $N$, which satisfies $\beta_{\eta,d,c}
\rightarrow 0$ when $d \rightarrow \infty$ and $\frac{c}{d}$ is kept
constant. By setting $\epsilon \rightarrow 1$ in \eqref{Rathi} and
using \eqref{Conditioning on Si}, we conclude that w.p. at least $1
- \frac{\beta_{\eta,d,c}}{\epsilon^2} + o(1)$, for $\mathcal{C}$
chosen randomly with uniform probability from
$\mathcal{C}^{\gamma}$,
\begin{equation}\label{inequality for Si}
    \frac{1}{N} \log S_{\eta N}^{\mathcal{C}} \geq \frac{1}{N} \log
    \overline{S}_{\eta N} + o(1)
\end{equation}

Define
\begin{eqnarray}
E_1 &\triangleq& -\max_{\gamma \leq \eta \leq \epsilon }
\left\{(1-\eta) h \left(\frac{\epsilon-\eta}{1-\eta}\right) +
\frac{1}{N}\log \overline{S}_{\eta
N}  \right\} \label{Average Exponent E1} \\
E_2 &\triangleq& -\max_{\substack{\gamma \leq \eta_2 \leq \eta_1
\leq \epsilon
\\ 0 \leq \beta \leq \eta_2 \\ \eta_1 + \eta_2 -\beta \leq \epsilon}} \left\{ (1-(\eta_1+\eta_2-\beta))h
\left(\frac{\epsilon-(\eta_1 + \eta_2 -\beta)}{1-(\eta_1 + \eta_2
-\beta)}\right) + \frac{1}{N}\log \overline{S}_{\eta_1 N,\eta_2
N,\beta N}  \right\} \label{Average Exponent E2}
\end{eqnarray}
then by combining \eqref{Pe^C}, \eqref{E_1'}, \eqref{E_2'},
\eqref{inequality for Sijk} and \eqref{inequality for Si}, we
obtain that, w.p. at least $1 -
\frac{\beta_{\eta,d,c}}{\epsilon^2} + o(1)$,
\begin{equation}\label{Pex exponential with average quantities}
    P_e^{\mathcal{C}}(\epsilon, N) \geq \frac{2}{r_{\epsilon N}+1}2^{-N
(E_1+o(1))}-\frac{2}{r_{\epsilon N}(r_{\epsilon N}+1)}2^{-N
(E_2+o(1))}
\end{equation}

As we are interested in the asymptotic behavior of $E_1$ and $E_2$
(and thus the exponential growth rate of the stopping set
distributions), we use \cite[Theorem 2]{Burshtein_Miller_IT04},
which asserts the following\footnote{Here we give the multivariate
version of the theorem with 3 variables; the theorem generalizes to
any number of variables.}:

Let $p(x,y,z)$ be a trivariate polynomial with non-negative
coefficients. Let $\alpha_1>0,\alpha_2>0$ and $\alpha_3>0$ be some
rational numbers and let $n_i$ be the series of all indices such
that $$\left[x^{\alpha_1 n_i}y^{\alpha_2 n_i}z^{\alpha_3 n_i}
\right] p(x,y,z)^{n_i} \neq 0$$ Then

\begin{equation}\label{BM Theorem 2}
    \lim_{i\rightarrow \infty} \frac{1}{n_i}\log \left[x^{\alpha_1 n_i}y^{\alpha_2 n_i}z^{\alpha_3 n_i}
\right] p(x,y,z)^{n_i}= \inf_{x>0,y>0,z>0} \log \left(
\frac{p(x,y,z)}{x^{\alpha_1}y^{\alpha_2}x^{\alpha_3}} \right)
\end{equation}
Using \eqref{Si}, \eqref{Sijk}, \eqref{Average Exponent E1},
\eqref{Average Exponent E2} and \eqref{BM Theorem 2} we obtain
\begin{eqnarray}
    E_1 &=& -h(\epsilon)-\max_{\gamma \leq \eta \leq \epsilon} \left\{ \epsilon h
\left(\frac{\eta}{\epsilon} \right) -c h(\eta) + \frac{c}{d}
\inf_{x>0} \log \left( \frac{1+\psi_2(x,d)}{x^{\eta d}} \right)
\right\} \label{E_1 final} \\
E_2 &=& -h(\epsilon)-\max_{\substack{\gamma \leq \eta_1 \leq \eta_2
\leq \epsilon \\ 0 \leq \beta \leq \eta_2
\\ 0 \leq \eta_1+\eta_2-\beta \leq \epsilon}} \left\{ \epsilon h\left(
\frac{\eta_1-\beta}{\epsilon}, \frac{\eta_2-\beta} {\epsilon},
\frac{\beta}{\epsilon} \right) - c h\left( \eta_1-\beta,
\eta_2-\beta, \beta \right) \right. \nonumber \\ && \left.
\hspace*{6cm} + \frac{c}{d} \inf_{x,y,z>0} \log \left(
\frac{B(x,y,z,d)}{x^{(\eta_1-\beta)d}y^{\beta d}z^{(\eta_2-\beta)d}}
\right) \right\} \nonumber 
\end{eqnarray}

If $E_2 \geq E_1$, we choose $r_{\epsilon N}=1$ in \eqref{Pex
exponential with average quantities}. In this case, taking the union
bound over all possible stopping sets yields an exponentially tight
bound. In the case that $E_2<E_1$, we use \eqref{Pex exponential
with average quantities} with $r_{\epsilon N}=\lfloor
2^{N(E_1-E_2+\alpha)}\rfloor$, where $\alpha>0$ can be made
arbitrarily small (hence, the non-negativity of
$\hat{P}_e^{\mathcal{C}}(\epsilon,N)$ in \eqref{PeCtag} is
established). Thus, we obtain the following upper bound on the error
exponent
\begin{eqnarray} \label{Upper bound on exponent}
  -\frac{1}{N} \log P_e^{\mathcal{C}} &<& -\max_{\gamma \leq \epsilon \leq 1}
\left\{\epsilon \log \delta + (1-\epsilon) \log (1-\delta) - \left\{ \begin{array}{l l} E & \gamma \leq \epsilon \leq 1-R \\
-h(\epsilon) & 1-R \leq \epsilon \leq 1 \end{array} \right.
\right\}   + o(1)
\nonumber \\ E &\triangleq& \left\{ \begin{array}{l l} E_1 & E_2 \geq E_1 \\
2E_1-E_2 & E_2<E_1 \end{array} \right.
\end{eqnarray}
This bound holds w.p. at least $1 -
\frac{\beta_{\eta_0,d,c}}{\epsilon^2} + o(1)$, where $\eta_0$ is
the maximizing value of $\eta$ in \eqref{E_1 final}.

\section{Numerical Results}\label{Numerical Results}
In this section, we compare our upper bound on the error exponent of
the BEC with previously-known lower bounds. These bounds were
derived in \cite[Theorems 8,12]{Burshtein_Miller_IT04}; one of these
bounds applies for iterative decoding, while the other applies for
ML decoding.

\begin{figure}[here!]
\begin{center}
\includegraphics[scale=0.9]{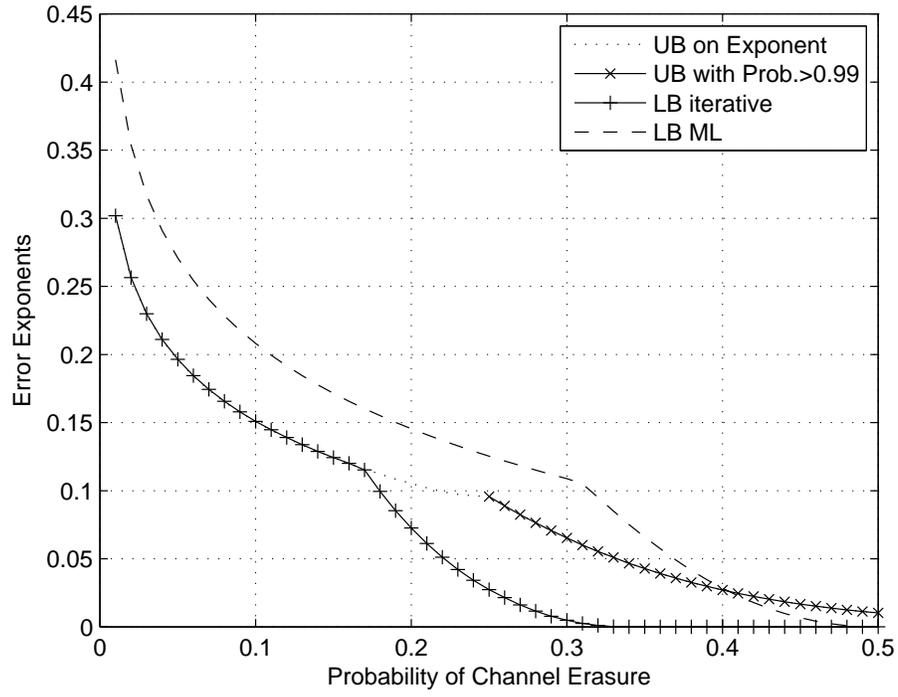}
\end{center}
\caption{Error exponents for the regular (4,8) LDPC ensemble.}
\label{Error Exponents for (4,8) ensemble}
\end{figure}

\begin{figure}[here!]
\begin{center}
\includegraphics[scale=0.9]{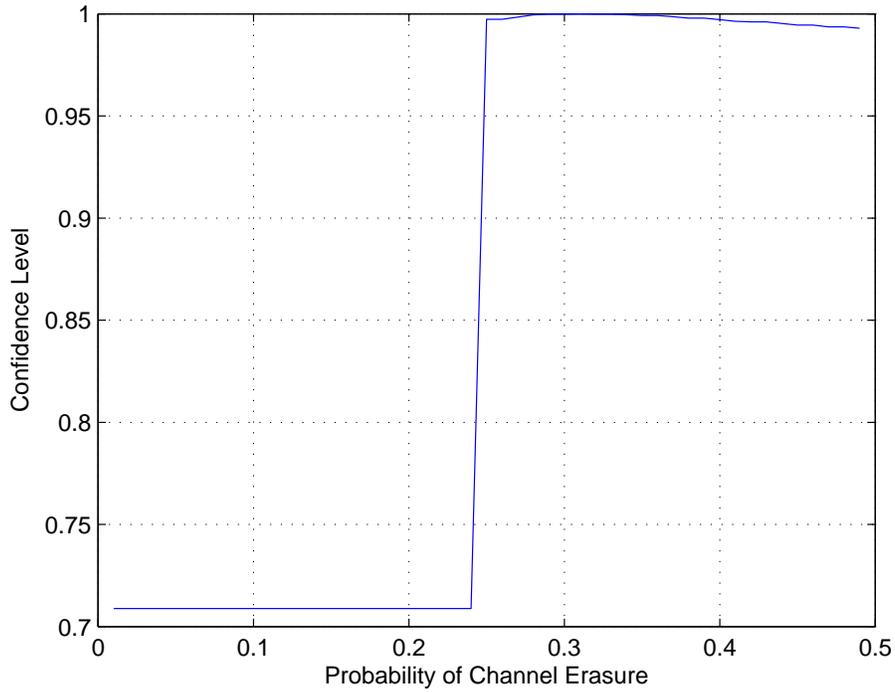}
\end{center}
\caption{Confidence level bound for the regular (4,8) LDPC
ensemble.} \label{Confidence Level for (4,8) ensemble}
\end{figure}

In Figure \ref{Error Exponents for (4,8) ensemble} we exemplify
our bound for the regular $(4,8)$ LDPC ensemble. Recalling that
the bound applies with a certain probability, we have marked the
plot where the bound has a confidence level above $99\%$. We note
that the entire plot of the upper bound is true w.p. at least
$70\%$.

Figure \ref{Confidence Level for (4,8) ensemble} shows the
confidence level bound from \eqref{Rathi} which corresponds to the
upper bound plot in Figure \ref{Error Exponents for (4,8) ensemble}.
Looking back at Figure \ref{Error Exponents for (4,8) ensemble} for
low values of $\delta$, the upper bound on the exponent coincides
with the two lower bounds from~\cite[Theorems
6,8]{Burshtein_Miller_IT04}. That is, our results indicate that in
the region $\delta\in [0,0.17]$, the bound on the error exponent of
the expurgated ensemble in~\cite[Theorem 6]{Burshtein_Miller_IT04},
which coincides with the bound in~\cite[Theorem
8]{Burshtein_Miller_IT04} in this region, is tight. Similarly, for
the (3, 6) ensemble and $\delta \in [0,0.26]$, the lower bound on
the error exponent of the expurgated ensemble in [7, Theorems 6]
(which coincides with the lower bound in [7, Theorem 8] in this
region) is tight\footnote{We note that these lower bounds, as
depicted in \cite[Figure 3]{Burshtein_Miller_IT04} do not coincide
with each other in this $\delta$ region due to a numerical
inaccuracy.}.

Focussing on higher values of $\delta$ where the confidence level is
higher, comparison of our upper bound with the lower bound on the ML
decoding exponent reveals that there is a gap in performance between
iterative and ML decoders, at least for most codes in the ensemble.

\section{Conclusion and Further Research}\label{Conclusion}
We have derived an upper bound on the error exponent of LDPC codes
transmitted over the BEC. The upper bound relies on Dawson's
inequality and holds with a certain confidence level. It was
demonstrated that for some values of the channel erasure
probability there is a gap between our upper bound and some
previously reported lower bounds.

Continued research could focus on extending our results to irregular
ensembles of LDPC codes. This requires to extend the results of
\cite{Rathi2006}, regarding concentration of stopping sets, to
irregular codes. Another possible avenue is to try and bridge the
gap between the lower and upper bounds; with the asymptotic decoding
threshold for the $(4,8)$ ensemble at about $0.38$, there is still
room for improvement in the bounds.


\section*{Acknowledgment} The authors wish to thank Igal Sason for
pointing out the improvement that was implemented in Equation
\eqref{Probability of erasure, true indicator}, and for
stimulating discussions.

%

\newpage
\section*{\center{\LARGE{Appendices}}}

\appendices

\section{Proof of \eqref{Dawson's generalized inequality}} \label{Proof of Dawson's
generalized}

Given the events $A_1,\dots,A_M$ define the set $B_s$,
$s=1,\dots,M$ as the set of points in $\bigcup_{i=1}^M A_i$
contained in exactly $s$ sets. We thus have
\begin{eqnarray}
  \sum_{k=1}^M k \mathbf{1}_{\{B_k\}} &=& \sum_{k=1}^M \mathbf{1}_{\{A_k\}} = S_1 \label{S_1} \\
  \sum_{k=2}^M \binom{k}{2} \mathbf{1}_{\{B_k\}} &=& \sum_{k=1}^M \sum_{i=1}^{k-1} \mathbf{1}_{\{A_k\}}\mathbf{1}_{\{A_i\}} = S_2 \label{S_2}
\end{eqnarray}
We will find a lower bound for
\begin{equation}\label{V}
    V=\mathbf{1}_{\left\{ \bigcup_{i=1}^M A_i \right\}}=\sum_{k=1}^M \mathbf{1}_{\{B_k\}}
\end{equation}
First, fix the value of $r$. Solving \eqref{S_1} and \eqref{S_2}
to isolate $\mathbf{1}_{\{B_r\}}$ and $\mathbf{1}_{\{B_{r+1}\}}$
we get
\begin{eqnarray}
  \mathbf{1}_{\{B_r\}} &=& S_1 - \frac{2S_2}{r} - \mathbf{1}_{\{B_1\}} - \sum_{\substack{k=2 \\ k \neq r}}^M \mathbf{1}_{\{B_k\}} \frac{k(r+1-k)}{r} \label{Indicator on B_r}\\
  \mathbf{1}_{\{B_{r+1}\}} &=&
\mathbf{1}_{\{B_1\}} \frac{r-1}{r+1} +
\frac{2S_2}{r+1}-S_1\frac{r-1}{r+1}- \sum_{\substack{k=2 \\ k \neq
r+1 }}^M \mathbf{1}_{\{B_k\}} \frac{k(k-r)}{r+1} \label{Indicator
on B_r+1}
\end{eqnarray}
Substituting \eqref{Indicator on B_r} and \eqref{Indicator on
B_r+1} into \eqref{V} we get
\begin{equation}\label{V alternative}
    V-\frac{2S_1}{r+1}+\frac{2S_2}{r(r+1)}=\frac{r-1}{r+1}\mathbf{1}_{\{B_1\}}+\sum_{k=2}^M
\mathbf{1}_{\{B_k\}}\frac{(r-k)(r-k+1)}{r(r+1)}
\end{equation}
Note that the RHS of \eqref{V alternative} contains only
non-negative elements. Thus, if the RHS of \eqref{V alternative}
is replaced by zero, we obtain the inequality
\begin{equation*}\label{final}
    V \geq \frac{2}{r+1}S_1 - \frac{2}{r(r+1)}S_2
\end{equation*}
which is the desired result.

\section{Confidence Interval of Stopping Set Distribution}\label{Confidence Interval of Stopping Set Distribution}
Rathi \cite{Rathi2006} has obtained a result asserting the
concentration of the stopping set distribution. To state his
result, we introduce some notation.

\begin{itemize}
\item
    Denote $\beta(x)\triangleq 1+\psi_2(x,d)$, where $\psi$ is
defined in \eqref{little psi}.
\item
The equation
\begin{equation*}\label{no}
    x\frac{(1+x)^{d-1}-1}{\beta(x)}=\eta
\end{equation*}
has a single real positive solution; denote this solution by
$x_{\eta}$. \item Define $a_{\beta}(x)\triangleq
\frac{x}{\beta(x)}\frac{\text{d}\beta(x)}{\text{d}x}$ and
$b_{\beta}(x)\triangleq x \frac{\text{d}a_{\beta}(x)}{\text{d}x}$
\item
    Let $\underline{x}=(x_1,x_2,x_3)$. For a multivariate function
    $f(\underline{x})$, denote $a_f(\underline{x})$ to be a 3-element
    vector whose elements are $a_{f(i)}=\left(
    \frac{x_i}{f}\frac{\partial f}{\partial x_i} \right)$.
    Let $C_f(\underline{x})$ denote a $3\times 3$ matrix whose
    elements are given by $C_{f(i,j)}=x_j\frac{\partial
    a_{f(i)}}{\partial x_j}=C_{f(j,i)}$.
\end{itemize}
The concentration result is as follows. The number of stopping
sets $S_{\eta N}^{\mathcal{C}}$ in a randomly selected code
$\mathcal{C}$ satisfies
\begin{equation}\label{rathi's result}
    \Pr \left( 1-\epsilon \leq \frac {S_{\eta N}^{\mathcal{C}}}
    {\overline{S}_{\eta N}} \leq 1 + \epsilon\right) \geq
    1 - \frac{\beta_{\eta,d,c}}{\epsilon^2} + o(1)
\end{equation}
where
\begin{eqnarray*}
  \beta_{\eta,d,c} &=& \frac{b_{\beta}(x_{\eta})\sqrt{d}\eta (1-\eta) \sigma_c(\eta^2)}
  {\sqrt{|C_{\tilde{B}}(x_{\eta},x_{\eta}^2,x_{\eta})|(\eta^2 (1-\eta)^2 - (c-1)\sigma^2_c(\eta^2)) }} -1 \nonumber \\
  \sigma^2_c(\eta^2) &=& \frac{1}{cd |(-1,1,-1) \cdot C_{\tilde{B}}(x_{\eta},x_{\eta}^2,x_{\eta})^{-1}\cdot (-1,1,-1)^T  |
  } \nonumber \\
  \tilde{B}(\underline{x})&\triangleq&B(x_1,x_2,x_3,d)
\end{eqnarray*}
and $B(\cdot,\cdot,\cdot,d)$ is defined in \eqref{P_(s,2)(i)}.

\end{document}